\documentclass[11pt]{article}

\usepackage[letterpaper,margin=1in]{geometry}

\usepackage{amsmath,amssymb}
\usepackage{graphicx}
\usepackage{url}
\usepackage[round]{natbib}
\usepackage{xcolor}
\usepackage{longtable}
\usepackage{booktabs}
\usepackage{siunitx}

\usepackage{authblk}

\usepackage[format=hang]{caption}
\definecolor{linkblue}{rgb}{0.0,0.0,0.6}
\usepackage[colorlinks=true,linkcolor=linkblue,citecolor=linkblue,urlcolor=linkblue]{hyperref}

\providecommand{\figureref}[1]{Figure~\ref{#1}}
\providecommand{\tableref}[1]{Table~\ref{#1}}
\providecommand{\sectionref}[1]{Section~\ref{#1}}

\makeatletter
\providecommand{\floatconts}[3]{%
  \def\@tempa{table}%
  \ifx\@captype\@tempa
    #2\label{#1}\par{\centering #3\par}%
  \else
    {\centering #3\par}#2\label{#1}%
  \fi}
\makeatother

\newenvironment{keywords}{\par\medskip\noindent\textbf{Keywords:}~}{\par}

\newcommand{\audiosupp}{on our demo page}

\title{Can Music Generation Models Tap Along?\\
Evaluating Generative Approaches to Beat Tracking}

\author[1,2]{Kun Fang}
\author[3,4]{Ziyu Wang}
\author[1,2]{Ichiro Fujinaga}
\affil[1]{Schulich School of Music, McGill University, Montr\'{e}al, QC, Canada}
\affil[2]{Centre for Interdisciplinary Research in Music Media and Technology (CIRMMT), Montr\'{e}al, Canada}
\affil[3]{Courant Institute, New York University, New York, USA}
\affil[4]{Mohamed bin Zayed University of Artificial Intelligence (MBZUAI), Abu Dhabi, UAE}

\date{}

\begin{document}

\maketitle

\begin{abstract}
Existing studies have shown that music generative models can generate synchronized tracks that play alongside an input song. Can we extend this capability to develop a generative approach to beat tracking? To answer this question, we equip a pretrained MusicGen model with an audio-conditioning adapter to generate percussion tracks synchronized with the input music, from which we extract the beat positions. We train separate models with percussion targets of varying rhythmic complexity and compare them with two other generative models and conventional beat trackers. On GTZAN, our model achieves performance close to state-of-the-art beat trackers. Our model also demonstrates greater robustness: it shows the smallest drop in F-measure when drums are removed from the input at inference time. In certain genres, such as metal and disco, our model is also less likely to be misled by strong onsets or to miss beats with weak onsets. These results suggest that percussion generation has the potential to become a viable approach to beat tracking, and that its output is not merely a sequence of beat positions but an audible percussion track.

\begin{keywords}
beat tracking, evaluation on music generation models, percussion generation
\end{keywords}
\noindent\textbf{Demo:}~\url{https://kunfang98927.github.io/generative-beat-tracking/}
\end{abstract}

\section{Introduction}
\label{sec:introduction}

When listening to music, people often spontaneously synchronize their movements with the perceived beat, such as by nodding or tapping along, or, in the case of musicians, improvising an accompanying rhythmic part \citep{repp2013sensorimotor}. In these behaviors, the perceived beat is expressed through coordinated movement or musical performance rather than explicit timestamps. In contrast, conventional beat tracking models directly estimate beat positions from audio \citep{bock2016madmom,foscarin2024beatthis}. While effective, such models may learn statistically predictive features without necessarily capturing the aspects of rhythmic perception involved in human beat tracking.

Recent advances in music generation have enabled models to generate new tracks (e.g., a drum part) that accompany an input musical piece \citep{wu2022jukedrummer,rouard2025musicgenstem,strano2025stage}. Generating such accompaniment requires models to capture aspects of the underlying rhythmic structure, even without explicitly predicting beat positions. Much like human musicians, they can express the beat through musical performance. This motivates a new question: can beat tracking be approached as a generative task rather than a direct prediction problem?

\begin{figure}[!t]
\floatconts
  {fig:intro}
  {\caption{Overall pipeline. The music generative model generates a percussion track from the song; we compare three targets and train one model per target. A lightweight beat tracker then extracts the beat positions.}}
  {\includegraphics[width=\linewidth]{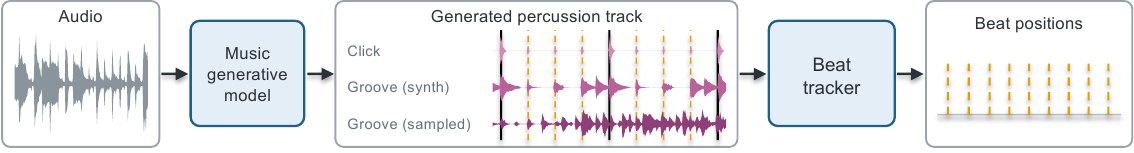}}
\end{figure}

We explore this generative approach by adapting a pretrained MusicGen model~\citep{copet2023musicgen} to generate percussion accompaniment conditioned on input music. We consider percussion targets of varying rhythmic complexity, from simple clicks to more elaborate drum grooves, and use a lightweight beat tracker to recover beat positions from the generated audio (\figureref{fig:intro}). We compare our approach with other generative methods and conventional beat trackers in terms of overall accuracy, and further investigate whether they can track beats beyond obvious acoustic cues. Specifically, we evaluate their performance when drums are removed and examine how onset strength relates to beat tracking errors.

Our experiments show that generative beat tracking achieves accuracy close to the state of the art, with our model reaching an F-measure of approximately 0.85 on GTZAN~\citep{tzanetakis2002gtzan} (\sectionref{sec:results}). We also find that our approach is more robust than the baselines in two ways. First, it shows the smallest performance drop when drums are removed from the input music (\sectionref{sec:drums}). Second, in certain genres, such as metal and disco, it is less likely to make incorrect beat predictions at peaks in onset strength or miss beats where onset strength is low (\sectionref{sec:onset-dep}). Audio examples of the generated percussion tracks are provided \audiosupp.
\section{Related work}
\label{sec:related}

\paragraph{Beat tracking.} 

Current beat tracking models map audio directly to beat positions, whether they are based on RNNs \citep{bock2016joint}, TCNs \citep{davies2019tcn,bock2020deconstruct}, or transformers \citep{hung2022spectnt,zhao2022beattransformer,foscarin2024beatthis}. More recently, diffusion models have also been explored for beat tracking \citep{foscarin2026masked}, but they still generate beat labels rather than audio.

\paragraph{Beat information in generative models.} Some studies use probing to read beat information from the hidden states of frozen generative models: \citet{wei2024syntheory} find that the hidden states of Jukebox and MusicGen contain tempo and time signature, \citet{zhao2026musicrfm} obtain similar results on MusicGen, and \citet{kanatas2026layer} read out beats from the hidden states of MusicGen. These findings suggest that generative models encode beat-related information, but do not establish whether this information can be used to generate rhythmically synchronized audio.

\paragraph{Conditional music generation.} Other work adds more precise control to generative models. MusicGen \citep{copet2023musicgen} originally generates music from text or melody, and later work has added conditions such as rhythm, chords, and drum tracks \citep{lan2024musicongen,lin2024cocomulla}. Most of these works give rhythm or drums to the model as input, so the model does not need to find the beat itself. Closer to this paper is work that takes existing audio as the condition and generates a new track that follows it, such as JukeDrummer \citep{wu2022jukedrummer}, MusicGen-Stem \citep{rouard2025musicgenstem}, and STAGE \citep{strano2025stage}. These approaches aim to generate musically consistent tracks, and typically evaluate beat alignment by comparing beats extracted from the generated and original tracks. In contrast, our training focuses on generating percussion with clear rhythmic patterns, and we evaluate the results against human annotations to assess beat-tracking accuracy.

\section{Methods}
\label{sec:methods}

We train a generative model on paired song audio and target percussion tracks synthesized from beat annotations. Given a 30-second music excerpt, the model generates a synchronized percussion track, from which a lightweight beat tracker extracts beat positions.

\subsection{Target percussion tracks}
\label{sec:targets}

We consider three percussion targets synthesized from the same beat annotations, varying in rhythmic complexity and timbre (shown in \figureref{fig:intro}). Simpler targets represent beat positions more explicitly, while more elaborate drum patterns may better resemble the music encountered during MusicGen's pretraining. We train a separate generative model for each target to investigate how these differences affect beat tracking performance.

\begin{enumerate}
\item \emph{Click} is the simplest target, representing each beat with a single percussion onset. We use a kick drum on downbeats and a hi-hat on other beats, producing a metronome-like pattern.
\item \emph{Groove (synth)} uses rule-based rhythmic patterns with synthesized percussion instruments, including kick, snare, hi-hat, cowbell, and conga. We define 14 genre-specific patterns, assigning one fixed pattern to each genre (\audiosupp).
\item \emph{Groove (sampled)} uses human-performed drum MIDI recordings from the Groove MIDI Dataset \citep{gillick2019groove}, selected by genre and tempo. We align the MIDI performances to annotated beats bar by bar and render them using drum samples from AVL Drumkits \citep{avldrumkits} (\audiosupp).
\end{enumerate}
Genre labels are taken from the dataset and used only to construct training targets; at inference, the model generates percussion conditioned solely on the input audio, without genre labels.

\subsection{Percussion generation model}
\label{sec:model}

\paragraph{Architecture.} MusicGen \citep{copet2023musicgen} is an autoregressive model that predicts EnCodec tokens \citep{defossez2022encodec} frame by frame, which are then decoded into audio. We introduce an audio-conditioning adapter connecting two pretrained MusicGen models (\figureref{fig:adapter}). The input song is encoded into EnCodec tokens $X_{1:T}$ and passed through the first MusicGen to extract frame-level features from each layer. These features are aggregated using a learnable weighted sum, followed by a stack of dilated convolutions that incorporates approximately 5 seconds of surrounding context. The resulting context $C_{1:T}$ is the condition of the second MusicGen, which autoregressively generates percussion EnCodec tokens $Y_{1:T}$. The context is incorporated in two ways: through frame-wise addition to its decoder input and gated cross-attention added to each transformer layer \citep{alayrac2022flamingo}. Both MusicGen models remain frozen, and only the adapter is trained. We use MusicGen-large to encode the input song and MusicGen-small to generate percussion, providing a good balance between performance and computational cost.\footnote{On the validation subset (\sectionref{sec:datasets}) and GTZAN, using MusicGen-small for music encoding reduces the F-measure by 0.013 and 0.003, respectively, while replacing the pretrained music encoder with an encoder trained from scratch leads to larger drops of 0.052 and 0.064. Pretrained MusicGen representations have also been shown to encode musical properties such as tempo \citep{wei2024syntheory}.}

\begin{figure}[!t]
\floatconts
  {fig:adapter}
  % {\caption{Our percussion generation model. The weights of the gray modules are fixed, and the orange modules are trained.}}
  {\caption{Our percussion generation model. $N$ and $N'$ are the numbers of layers of the two MusicGen models; $T$ is the number of frames. Gray modules are frozen, and orange modules are trained.}}
{\includegraphics[width=0.99\linewidth]{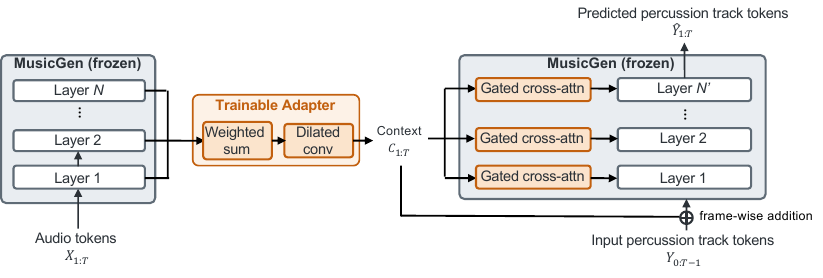}}
\end{figure}

\paragraph{Training and inference.} We adopt two training strategies to improve the rhythmic alignment of generated percussion with the input music. First, with probability 0.8, we replace the autoregressive context with a silence token to discourage the model from repeating previously generated rhythms without attending to the song.\footnote{Without this strategy, the F-measure on the validation subset drops to 0.31.} Second, with probability 0.5, we replace the input song with drum-removed audio obtained using HT Demucs \citep{rouard2023htdemucs}, encouraging the model to infer beats from other instruments  (see \sectionref{sec:drums}). We train each model for 30 epochs on 10-second excerpts using cross-entropy loss on the target EnCodec tokens, with a cosine learning rate schedule from $3\times10^{-4}$ to 0.

During validation, we use madmom to extract beats from the generated percussion tracks and select the checkpoint by the F-measure (70\,ms) on the validation subset. At inference, the model generates a 30-second track in one pass from the first 30 seconds of the song, using greedy decoding. To prevent the generated rhythm from drifting away from the song and merely continuing its own earlier rhythm, we use classifier-free guidance (CFG; \citealp{ho2022cfg}) with a coefficient of 1.5 at inference to strengthen the influence of the condition.

\subsection{Beat extraction}
\label{sec:extraction}

We train a lightweight beat tracker to extract beat positions from the generated percussion, since not all percussion onsets coincide with beats, and the generated audio may occasionally contain artifacts. Existing beat trackers are typically trained on full music and may not perform equally well on different types of percussion. To ensure a fair comparison across targets, we train a separate tracker for each using the same architecture and training procedure. Following \citet{bock2020deconstruct}, each tracker uses a temporal convolutional network (TCN) with three convolutional layers and 11 dilated convolution blocks to predict frame-level beat probabilities from the log-mel spectrogram of 16\,kHz audio. A dynamic Bayesian network (DBN) decoder then extracts beat positions. Each tracker has approximately 143\,K parameters.

\section{Beat tracking evaluation}
\label{sec:evaluation}

\subsection{Datasets}
\label{sec:datasets}

We use the beat annotations and the train/validation split compiled by Beat This! \citep{foscarin2024beatthis}. The models are trained on the training split of Ballroom \citep{gouyon2006ballroom}, Hainsworth \citep{hainsworth2004particle}, RWC \citep{goto2002rwc}, and ASAP \citep{foscarin2020asap}. The TCN beat tracker of \sectionref{sec:extraction} is trained with the percussion tracks generated from 300 of these songs as its input, 75 from each dataset.

We evaluate all methods on two sets. The first is a fixed subset of the validation split with 124 songs (40 from Ballroom, 33 from Hainsworth, 21 from RWC, and 30 from ASAP; the songs from Ballroom and ASAP are randomly sampled, to make the datasets closer in size and to reduce computation). The second is the 999 songs of GTZAN \citep{tzanetakis2002gtzan}, which serves as a held-out test set.

\subsection{Compared methods}
\label{sec:compared}

To evaluate the approach of first generating a percussion track and then extracting beats, we compare it with two other audio generation models and with two traditional beat tracking models. \tableref{tab:gtzan} (\sectionref{sec:results}) summarizes all methods.

\paragraph{Stable Audio ControlNet.} While our main model uses autoregressive generation, as in much of the prior work on conditional music generation \citep{lin2024cocomulla,strano2025stage}, we also explore a diffusion-based alternative using Stable Audio Open 1.0 \citep{evans2024stableaudioopen}. Following ControlNet \citep{zhang2023controlnet}, we freeze the base model and train a conditioning branch copied from its first 12 layers. This branch takes the song latent as input and adds its layer-wise outputs to the corresponding layers of the base model. We use the same training songs and percussion targets as Groove (sampled), allowing us to compare the two generative approaches and their conditioning mechanisms.

\paragraph{STAGE.} STAGE \citep{strano2025stage} is an existing drum accompaniment generation model, which we use to examine how accurately the drum accompaniment generated by an existing model tracks the beat. Its training target is drum stems of human performances. Like our model it is based on MusicGen-small, but it fine-tunes all weights and prepends the tokens of the context audio to the sequence as the condition. We use the weights released by the authors and the default sampling settings, with the full mix of the song as input. Since it is trained on 10-second excerpts, we generate the 30-second track in segments, each time continuing for 5 seconds from the last 5 seconds generated so far.

We also train a TCN beat tracker as in \sectionref{sec:extraction} for each of Stable Audio ControlNet and STAGE, so that all generative models extract beats in the same way.

\paragraph{Traditional beat tracking models.} These methods extract beat positions directly from the song audio. Beat This! \citep{foscarin2024beatthis} is based on a Transformer, needs no post-processing, and is one of the current state-of-the-art methods; madmom \citep{bock2016madmom} is based on a recurrent neural network (RNN) and is a classic baseline. The training data of madmom include Ballroom and Hainsworth, so it may have seen some songs in the validation subset. None of the models has seen GTZAN.

\subsection{Evaluation metrics}
\label{sec:metrics}

All methods take the full mix of the song as input and use only the first 30 seconds of the song. We use mir\_eval \citep{raffel2014mireval} to compare the beat positions extracted by each method with the beat annotations. F-measure measures whether each beat falls near an annotation, and we report it with three tolerances: 70\,ms, 100\,ms, and 150\,ms. CMLt requires both the metrical level and the phase to be correct, while AMLt also allows double tempo, half tempo, and off-beat; CMLc and AMLc consider only the longest continuous segment of correct beats, to measure continuity. For the generated tracks, we also measure audio quality with the Production Quality of Audiobox Aesthetics (PQ, higher is better; \citealp{tjandra2025audiobox}). All results are computed per song and then averaged.

\subsection{Results}
\label{sec:results}

\tableref{tab:gtzan} shows the results of all methods on GTZAN. Our generative approaches can reach an accuracy close to that of traditional beat tracking models. The choice of target has almost no effect: from Click to the Groove targets, which are closer to real performance, there is no visible difference in F-measure. For Stable Audio ControlNet, one possible reason for its lower F-measure is temporal precision: each frame of its latent is about 46\,ms long, compared with 20\,ms for MusicGen. With relaxed tolerances, the ranking stays the same; Stable Audio ControlNet and STAGE improve the most in F-measure, which shows that many of their beats are not far from the annotations.

\begin{table}[!t]
\floatconts
  {tab:gtzan}
  {\caption{Results on GTZAN (full-mix input). Top: generative methods; bottom: traditional beat tracking models. F$_{x}$: F-measure with a $\pm x$\,ms tolerance. Params: trainable parameters, excluding the TCN beat tracker. PQ: Production Quality of Audiobox Aesthetics (1--10). Bold: best; underline: second best.}}
  {\footnotesize\setlength{\tabcolsep}{2.2pt}\renewcommand{\arraystretch}{0.92}
  \begin{tabular}{lrrcccccccc}
    \toprule
    Method & Training songs & Params & F$_{70}$ & F$_{100}$ & F$_{150}$ & CMLt & AMLt & CMLc & AMLc & PQ \\
    \midrule
    Ours (Click) & 1319 (49\,h) & 10.4\,M & 0.851 & 0.871 & 0.889 & 0.756 & 0.907 & 0.731 & 0.874 & 6.27 \\
    Ours (Groove synth) & 1319 (49\,h) & 10.4\,M & 0.850 & 0.866 & 0.882 & 0.736 & \underline{0.912} & 0.713 & \underline{0.881} & 6.88 \\
    Ours (Groove sampled) & 1319 (49\,h) & 10.4\,M & 0.849 & 0.863 & 0.878 & 0.727 & 0.907 & 0.704 & 0.877 & 5.93 \\
    Stable Audio ControlNet & 1319 (49\,h) & 561\,M & 0.691 & 0.728 & 0.772 & 0.553 & 0.753 & 0.512 & 0.687 & \textbf{7.54} \\
    STAGE & 214 (13\,h) & 420\,M & 0.335 & 0.443 & 0.609 & 0.244 & 0.482 & 0.201 & 0.387 & \underline{7.09} \\
    \midrule
    Beat This! & 4556 (167\,h) & 20.3\,M & \textbf{0.888} & \textbf{0.901} & \textbf{0.913} & \textbf{0.793} & 0.895 & \underline{0.751} & 0.834 & -- \\
    madmom & $\sim$1100 (12\,h) & 0.73\,M & \underline{0.870} & \underline{0.884} & \underline{0.898} & \underline{0.773} & \textbf{0.923} & \textbf{0.753} & \textbf{0.894} & -- \\
    \bottomrule
  \end{tabular}}
\end{table}

AMLt best shows the character of this approach. It allows double tempo, half tempo, and off-beat, so it mainly reflects whether the model keeps up with the song. Our models are below Beat This! in F-measure and CMLt but above it in AMLt; the same holds for AMLc. This shows that the generated tracks have good continuity: the models mostly keep up with the song, and the main errors lie in the metrical level or phase rather than in losing the song entirely. Our TCN beat tracker and madmom both use a DBN (Beat This! does not), which by itself favors continuity, and madmom still has a higher AMLt than our models; but when the same pipeline is applied to Stable Audio ControlNet and STAGE, AMLt is much lower, so the DBN is not enough to explain this result.

To show that the accuracy of our models does not depend too much on the TCN beat tracker, we ran two controls with Ours (Click) as an example. First, instead of reading the beats from the percussion track with any beat tracker, we apply onset peak-picking directly to the generated track (with parameters tuned only on the validation subset); the F-measure is still 0.747 (0.851 with the TCN beat tracker). Second, when the same TCN beat tracker reads the original song directly, the F-measure is only 0.373. Thus the beats are given mainly by the generated track, and the TCN beat tracker plays a corrective role.

In audio quality, Stable Audio ControlNet and STAGE have higher PQ, and our Groove (sampled) model has the lowest. This may be because we train only an adapter of 10.4\,M parameters and keep the generator itself frozen, and because this task requires the generated track both to keep up with the song frame by frame and to have realistic timbre, which is challenging in itself. By ear, the tracks of Stable Audio ControlNet and STAGE sound more like real performances, but they are often misaligned with the beat or add rather arbitrary fills. The supplementary material provides examples of the tracks generated by each method.

\section{Do models rely on drums and onsets?}
\label{sec:onset}

This section examines two common problems in beat tracking: \sectionref{sec:drums} removes the drums from the audio to see whether the methods rely too much on drums, and \sectionref{sec:onset-dep} examines to what extent the beats of each method depend on onset strength. All experiments in this section use the Groove (sampled) model.

\subsection{Robustness to drum removal}
\label{sec:drums}

We often observe that beat tracking models perform very well on music with clear drums but become clearly worse when the drums are weak or absent. Drums are the most direct cue to the beat, but without drums humans can still tap the beat correctly by relying on other cues, and a good beat tracking method should not rely too much on them either. To examine whether the methods can still find the beat when the drums disappear, at inference we replace the model input with audio from which the drums have been removed by HT Demucs, keeping all other settings unchanged; ASAP is solo piano and has no drums, so it is not separated. For a fair comparison, we also need to consider whether each method has seen drum-removed audio during training: the models reported in \sectionref{sec:results} and Stable Audio ControlNet were trained with drum-removal (DR) augmentation (\sectionref{sec:model}), whereas, to the best of our knowledge, Beat This! and madmom were not. We therefore additionally train a model without this augmentation, denoted Ours (w/o DR aug), which, like Beat This! and madmom, has not seen drum-removed audio.

\figureref{fig:drums} compares the scores of the methods before and after the drums are removed. Even Ours (w/o DR aug) is robust: in both F-measure and AMLt, it drops less than Beat This! and madmom. Although Ours (w/o DR aug) has a lower F-measure than these two models on the full mix, it catches up after the drums are removed: its F-measure is level with Beat This! and clearly above madmom; and its AMLt is level with madmom and clearly above Beat This!.

Our model may benefit from autoregressive generation. The model generates the track step by step, with each step based on what has already been generated, so when the rhythmic cues in the song weaken, it can still continue along the rhythm it has established. However, not all generative models have this benefit: Stable Audio ControlNet, which is based on a diffusion model, drops as much as madmom in F-measure even though it was trained with DR augmentation. It generates the whole audio at once, without this step-by-step continuation, which may be one of the reasons, but confirming this requires more experiments.

\begin{figure}[!t]
\floatconts
  {fig:drums}
  {\caption{F-measure (70\,ms, top) and AMLt (bottom) of each method before and after the drum track is removed. Filled markers are the original song and open markers the drum-removed audio; a shorter connecting line means a smaller drop and greater robustness. Numbers below the lines are the drops. Ours and Ours (w/o DR aug) both use the Groove (sampled) target; w/o DR aug denotes training without drum-removal augmentation.}}
  {\includegraphics[width=0.99\linewidth]{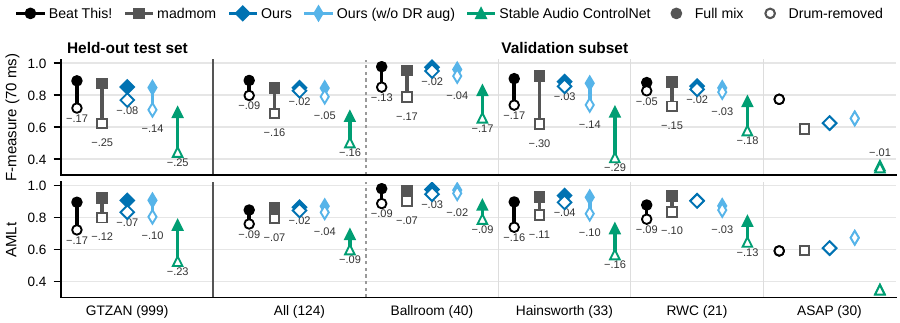}}
\end{figure}

\subsection{Dependence on onsets}
\label{sec:onset-dep}

\paragraph{Measures.}\label{sec:onset-measures}
We often observe that beat tracking models tend to place beats on loud onsets even when these are not beats (for example, when a percussive onset falls on the off-beat, the beats of a model may shift by half a beat as a whole; see \figureref{fig:onset-example}), and that they easily miss beats at soft positions. Previous studies have pointed out that soft onsets, sounds lacking transients, and very soft accompaniment are common difficulties for beat tracking, whereas soft onsets do not markedly increase the difficulty of perceiving the beat for humans \citep{grosche2010difficult,holzapfel2012selective}. A generative model may be less prone to such onset-dependent errors, as it needs to keep the generated audio coherent. We therefore examine how much the beats of each method depend on onsets.

We measure onset strength by spectral flux. Let $L(m, t)$ be the value of the log-mel spectrogram of the input audio in mel band $m$ and frame $t$. The onset strength of frame $t$ is
\begin{equation}
s(t) = \frac{f(t) - \mu}{\sigma}, \qquad f(t) = \sum_{m} \max\bigl( L(m, t) - L(m, t-1),\ 0 \bigr),
\end{equation}
where $\mu$ and $\sigma$ are the mean and standard deviation of $f(t)$ over the song, so that $s(t)$ indicates how strong or weak an onset is within this song.

This dependence can show up in two ways. \emph{First, the weaker the onset on a beat, the more easily the beat is missed.} We take the maximum of $s(t)$ within 30\,ms around each annotated beat as the onset strength of that beat, and divide all annotated beats into 5 groups by quantile. Let $B_k$ be the annotated beats of a song that fall in group $k$, and let $E$ be the beats extracted by the model. The miss rate of group $k$ is
\begin{equation}
M_k = 1 - \frac{\bigl|\{\, b \in B_k : \min_{e \in E} |b - e| \le 70\,\mathrm{ms} \,\}\bigr|}{|B_k|}.
\end{equation}

\emph{Second, a strong onset between beats is easily mistaken for a beat.} We take the local peaks of $s(t)$ above 0 as onsets and keep only those at least 140\,ms and at least 0.2 beats from the nearest annotated beat, which we call off-beat onsets (simply put, these are the onsets that are loud but not on the beat). They are likewise divided into 5 groups by quantile. Let $O_k$ be the off-beat onsets of a song that fall in group $k$. The false-positive rate of group $k$ is
\begin{equation}
F_k = \frac{\bigl|\{\, o \in O_k : \min_{e \in E} |o - e| \le 70\,\mathrm{ms} \,\}\bigr|}{|O_k|}.
\end{equation}

Both metrics are first computed within each song and then averaged over songs.

Octave errors (that is, tapping at double or half tempo) interfere with these two metrics: the extra beats of double tempo would be counted as false positives, and the beats skipped at half tempo would be counted as misses, but these errors may be unrelated to onset dependence. We therefore keep only the songs for which the tempos predicted by Beat This!, madmom, Ours, and Ours (w/o DR aug) all differ from the annotations by less than 10\%, and compare the methods on this same set of songs.

\paragraph{Results.}\label{sec:onset-results}
\figureref{fig:onset-genre} shows, by genre, how the miss rate and the false-positive rate vary with onset strength; lower is better for both. The miss rate answers the question \emph{how often does the model miss beats with weak onsets?}, and the false-positive rate answers \emph{how often does the model place beats on strong off-beat onsets?}

After the drums are removed, our Groove (sampled) model depends less on onsets than Beat This! and madmom, most clearly on metal and disco. On these two genres, our model has the lowest miss rate in every onset-strength group, and its false-positive rate stays low as off-beat onsets get stronger, whereas those of Beat This! and madmom rise steeply. Ours (w/o DR aug) lies between them. On rock and pop, madmom is clearly worse than the others, and on the other genres the methods are close. Classical music behaves differently: it has almost no drums, so removing them changes little, and the miss rate of all methods falls markedly as onsets get stronger.

\begin{figure}[!t]
\floatconts
  {fig:onset-genre}
  {\caption{Miss rate (top) and false-positive rate (bottom) as a function of onset strength for each genre of GTZAN on drum-removed audio; lower is better for both. Only the songs for which all methods have the correct tempo are used (458 in total).}}
  {\includegraphics[width=0.99\linewidth]{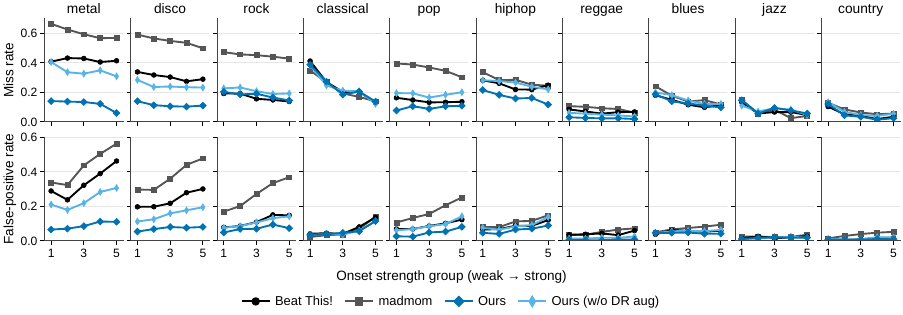}}
\end{figure}

\paragraph{Case study.} \figureref{fig:onset-example} shows one of the clearest examples of how removing the drums affects each method. On the full mix, the onsets on the annotated beats are almost all in group 4 or 5, and all methods place their beats correctly. After the drums are removed, most of these onsets drop to groups 1 to 3, while the onsets halfway between the beats mostly remain in group 5. Beat This! and madmom follow these strongest onsets: for almost the whole excerpt, their beats shift by half a beat onto the off-beat onsets (red), so they make both errors measured in \figureref{fig:onset-genre}. Ours and Ours (w/o DR aug), in contrast, keep their beats on the annotated beats, including those whose onsets are in group 1. This example also shows that drums and onsets are coupled. Drums usually produce the strongest onsets on the beats; once they are removed, strong onsets may fall between the beats and easily mislead models that rely heavily on onsets.

\begin{figure}[!t]
\floatconts
  {fig:onset-example}
  {\caption{Onset strength $s(t)$ and the beats of each method on gtzan\_disco\_00036 (11--16\,s), with the full mix on top and drum-removed audio below. Numbers give the onset strength group (1 = weakest, 5 = strongest).}}
  {\includegraphics[width=0.99\linewidth]{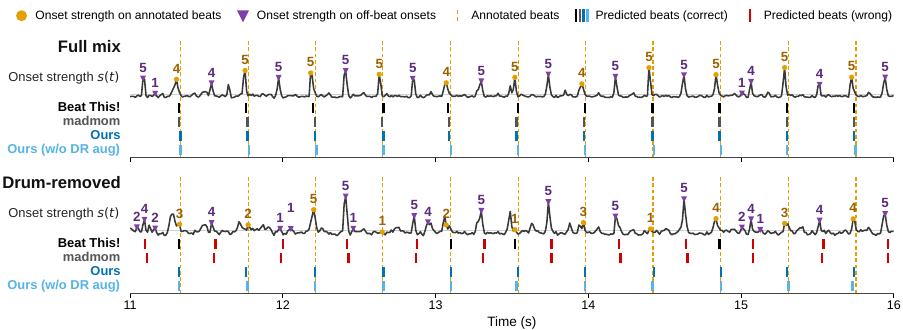}}
\end{figure}

\section{Discussion and conclusion}
\label{sec:discussion}

First, by incorporating a small adapter into a frozen MusicGen, we can extract beats from the generated percussion tracks with an accuracy comparable to that of traditional beat tracking models, while maintaining a consistent rhythm. This consistency may come from autoregressive generation, and it has a drawback and possibly a benefit: the model may merely continue its own rhythm without following the audio condition, but when the drums disappear, this continuation may also keep the beat predictions stable. Second, our model is less often misled by strong onsets that are not on the beat and less often misses beats with weak onsets. Such an analysis of onset dependence can complement the commonly used beat tracking metrics: metrics such as F-measure, CMLt, and AMLt only check whether the beats of a model align with the annotations, so a model may score highly even if it simply follows onsets.

This study has several limitations. The training data of the generative models are not entirely the same, so this is a comparison of existing methods rather than of model architectures. Drum-removed audio comes from source separation and is not equivalent to music that naturally has no drums. Audio quality was assessed only preliminarily, without a listening test. In conclusion, once trained with percussion tracks created from beat annotations as targets, a generative model can follow a song accurately enough, while also producing an audible percussion track.

\bibliographystyle{plainnat}
\bibliography{ref}

\end{document}